# Scanning Plasmonic Color Display

*Xiaoyang Duan,[1,2] and Na Liu[\*,1,2]*

[1]Max Planck Institute for Intelligent Systems, Heisenbergstrasse 3, 70569 Stuttgart, Germany.
[2]Kirchhoff Institute for Physics, University of Heidelberg, Im Neuenheimer Feld 227, 69120 Heidelberg, Germany.
[\*]E-mail: na.liu@kip.uni-heidelberg.de

Control over plasmonic colors on the nanoscale is of great interest for high-resolution display, imaging, and information encryption applications. However, so far very limited schemes have been attempted for dynamic plasmonic color generation. In this Letter, we demonstrate a scanning plasmonic color generation scheme, in which subwavelength plasmonic pixels can be laterally switched on/off through directional hydrogenation/dehydrogenation of a magnesium screen. We show several dynamic plasmonic color displays with different scanning functions by varying the number and geometries of the palladium gates, where hydrogen enters the scanning screens. In particular, we employ the scanning effects to create a dynamic plasmonic quick response code. The information cannot be decrypted by varying the polarization states of light or by accessing the physical features. Rather, it can only be read out using hydrogen as decoding key. Our work advances the established design concepts for plasmonic color printing as well as provides insights into the development of optical information storage and anticounterfeiting features.

Color generation has been a perpetual pursuit through art exploration and scientific advancement along the history of humankind. Compared to conventional coloration methods based on pigmentary, structural colors possess many advantages, such as high spatial resolution, material simplicity, environmental friendliness, and long-term durability.[1–11] In particular, structural color generation using plasmonic nanoparticles as subwavelength pixels has recently gained tremendous momentum. Plasmonic colors emerge from resonant interactions between light and metal nanostructures, resulting in collective oscillations of the conduction electrons, *i.e.*, the so-called localized surface plasmons.[12] With the drive towards practical applications, a variety of potential prototype devices have been implemented for color printing,[3–7] colorimetric sensing,[13,14] color filtering[15–18] as well as anticounterfeiting and authentication.[19–21] However, the bottleneck of plasmonic color generation lies in its static nature. To this end, several approaches have been suggested and successfully employed for plasmonic color tuning, including polarization control,[15–19] alignment of liquid crystals,[22–24] mechanical strain,[25,26] integration of phase-transition materials[27–32] and so forth.[33–36] Nevertheless, dynamic plasmonic color control so far has been limited to very primitive schemes. Only simultaneous tuning or switching on/off plasmonic colors has been shown.[27–35] Endowing plasmonic pixels with versatile and rigorous dynamics is essential to enrich the plasmonic color generation toolbox as well as to further promote a broader range of applications.

In this Letter, we demonstrate scanning plasmonic color displays, taking inspiration from macroscopic scanning devices. Scanning control over subwavelength plasmonic pixels, *i.e.*, aluminium (Al) nanoparticles is enabled by introducing an underlying magnesium (Mg) layer as scanning screen. Such a microscopic screen can be refreshed by laterally erasing or restoring plasmonic colors, when the Mg layer is transformed between the metal and dielectric (magnesium hydride, $MgH_2$) states through directional hydrogenation/dehydrogenation. We show several dynamic plasmonic color displays with versatile scanning effects. The potential of this scheme for information encryption is also demonstrated using a scanning plasmonic quick response (QR) code, which contains highly secure information.

Figure 1a illustrates the working principle of the scanning plasmonic color display. The microscopic scanning screen is a titanium (Ti, 3 nm)/Mg (30 nm) layer with dimensions of 15 × 15 μm$^2$, which resides on a silicon dioxide substrate. To impart the scanning characteristics, the left side of the screen is in contact with a palladium (Pd) strip, which works as a gate for hydrogen loading or unloading. The Ti layer helps to release the mechanical stress from volume expansions of Mg and Pd. It also plays an important role as spacer to prevent Mg and Pd from alloying.[21] Al nanoparticles, *i.e.*, the plasmonic pixels, are arranged on top of the scanning screen spaced by an aluminium oxide (Al$_2$O$_3$, 20 nm) layer. The Al particle diameter and interparticle gap are defined as $D$ and $g$, respectively. Upon hydrogen loading, hydrogenation of Mg starts from the Pd gate such that the plasmonic pixels are laterally scanned, following the hydrogen diffusion direction. During the process, the scanning screen transits from a mirror (Mg) to a transparent spacer (MgH$_2$). This process is reversible through dehydrogenation using oxygen.[21,27–29] Figure 1b presents a scanning electron microscopy (SEM) image of the palette (left-bottom corner), in which the different functional layers are clearly visible.

We first investigate the lateral diffusion process of hydrogen in Mg (see Figures 2a–2c), which governs the scanning function of the dynamic plasmonic color display. As the top and side surfaces of the Mg layer are protected by its native oxide (MgO), hydrogen can only enter Mg via the Pd gate, which catalyzes the dissociation of hydrogen molecules into atoms.[37,38] In order to characterize the diffusion front mobility, optical hydrogenography is utilized to *in situ* record the optical reflection (OR) images of the diffusion process at hydrogen concentration of 20% and 80 °C (see Figure 2a). The sharp color change at the diffusion front arising from the Mg to MgH$_2$ transition makes it possible to accurately monitor the time-dependent diffusion characteristics. The experimental data (red curve) in Figure 2c reveal a typical diffusive process following a nucleation step. More specifically, the square of the front position $x^2$ (see Figure 2b), is proportional to time $t$, after a short nucleation time, $t_0$. The front mobility $K$ is defined as

$$K = \frac{x^2}{t - t_0} \qquad (1)$$

which has the same dimension ($m^2 s^{-1}$) as a diffusion coefficient. Fitting of the experimental curve gives rise to $K = 0.949 \times 10^{-8}$ $m^2 s^{-1}$ and $t_0 = 326$ s (see the black-dashed curve in Figure 2c). This indicates the absence of blocking effects for lateral hydrogen diffusion, which is in sharp contrast to the out-of-plane diffusion.[39,40]

The process is reversible through dehydrogenation of $MgH_2$ using oxygen (20% at 80 °C) as shown in Figure 2d. The oxidative dehydrogenation involves binding of oxygen with the desorbed hydrogen atoms from $MgH_2$. This avoids a buildup of hydrogen at the Pd surface, thus facilitating hydrogen desorption.[38] It is noteworthy that the dehydrogenation process (the $MgH_2$ to Mg transition) also starts from the Pd gate, therefore giving rise to the same scanning direction (see Figure 2e) as that in the hydrogenation process. The top and side surfaces of the Mg screen are covered by the $Al_2O_3$ spacer and MgO, respectively, while its bottom surface resides on the substrate. Therefore, during dehydrogenation the desorbed hydrogen atoms from $MgH_2$ can only bind oxygen at the Pd gate. In addition, the high hydrogen concentration gradient across the gate region facilitates hydrogen desorption at the Pd gate, where $MgH_2$ is transformed into Mg. The measured time-dependent diffusion characteristics are presented by a blue curve in Figure 2f. Fitting of the experimental curve gives rise to $K = 1.037 \times 10^{-8}$ $m^2 s^{-1}$ and $t_0 = 2398$ s (see the black-dashed curve in Figure 2f).

Next, color generation from the Al nanoparticles on such a scanning screen is examined. Figures 3a and 3b present the reflection bright-field microscopy images of the palette, when the screen is in the Mg and $MgH_2$ states, respectively. When it is in the Mg state (color state, see Figure 3a), the palette exhibits a wealth of brilliant colors. In this case, the screening screen serves as a back reflector. The Al nanoparticle, the $Al_2O_3$ spacer, and the Mg layer construct a particle-on-mirror geometry[41]. The optical spectra of five representative color squares (i-v) from the palette are shown in Figure 3c. When the interparticle gap ($g$) is fixed, the particle diameter ($D$) increase leads to the red-shift of the resonance dip. The experimental and simulated reflectance spectra agree well. The simulated electric-field distribution at the resonance position (565 nm) of color square iv confirms the strong interactions between the Al particle plasmons and the Mg mirror, generating electromagnetic fields localized in the $Al_2O_3$ spacer. After hydrogenation, the scanning screen is in the $MgH_2$ state (plain state, see Figure 3b). The brilliant colors in the palette diminish. The experimental spectra of the selected color squares (i–v) exhibit nearly featureless profiles in the wavelength range of interest as shown in Figure 3d. In the absence of the mirror, the dipolar plasmon excitation of the Al nanoparticle results in a resonance peak, shifting to the red when the particle diameter ($D$) increases. The switching between the color and plain states of the palette is reversible through hydrogenation and dehydrogenation.

The interesting ability to laterally erase and restore plasmonic colors enables the realization of scanning plasmonic color displays. As a demonstration, a Sichuan opera facial mask has been employed as blueprint to design a dynamic plasmonic color display. Face-changing is an important subgenre of Chinese Sichuan opera. The SEM images in Figure 4a illustrate the details of the microprint. Figure 4b shows the performance of the scanning plasmonic display and the accompanying movie can be found in Supporting Movie 1. Upon hydrogen loading, hydrogenation starts from the Pd gate on the left. Absorption of hydrogen in Mg leads to a uniform diffusion front, scanning the microprint from the left to the right (see the red arrows). Along the diffusion route, the plasmonic colors are switched off laterally. Upon oxygen loading, dehydrogenation also starts from the Pd gate. Desorption of hydrogen in $MgH_2$ leads to the formation of Mg, therefore laterally switching on the plasmonic colors and restoring the color microprint from the left to the right (see the blue arrows).

Our scheme can be extended to generate a variety of scanning effects by tailoring the Pd gates, where hydrogen enters Mg. For instance, when a Pd disk is placed in the center of a firework display, the isotropic diffusion of hydrogen in Mg leads to a radially propagating effect as shown in Figure 4c (see also Supporting Movie 2). The detailed schematic of the firework display can be found in Supporting Figure S1. Inclusion of several Pd gates and variation of their positions add extra degrees of freedom to manipulate the scanning effects. Other firework display examples can be found in Supporting Figure S2 (see also Supporting Movie 2). Figure 4d shows a peacock display, in which the Pd stripe is curved along the tail of the peacock (see Supporting Figure S3 for the detailed schematic). Color scanning along the tail feathers is nicely visualized and the accompanying movie can be found in Supporting Movie 3.

The scanning functions enable novel information encryption and cryptography concepts using plasmonic colors. Traditional anticounterfeiting features with plasmonic nanostructures often utilize polarization control to switch between two pre-defined images.[17,19] However, this is insufficient to securely protect information against piracy, because the information can be straightforwardly decrypted by simply varying the polarization states. To deter forgery, the protected information should not be easily accessible through physical features. Figure 5a presents our design and working principle of information encryption as a proof-of-concept experiment. We have created two QR codes, QR1 and QR2, respectively. '0' in QR1 is represented using two bright colors, (*D*100, *g*150) and (*D*114, *g*100), which are selected from the palette in Figure 3. The distinction between the two bright colors is that after hydrogenation (*D*100, *g*150) is transformed to dark color (therefore '1'), whereas (*D*114, *g*100) remains bright (therefore '0') as shown in Figure 5a (also see Figure 3). '1' in QR1 is represented using two dark colors, (*D*060, *g*175) and (*D*080, *g*100) from the palette. The difference between (*D*060, *g*175) and (*D*080, *g*100) is that after hydrogenation the former is transformed to bright color (therefore '0'), whereas the latter remains dark (therefore '1'). In this context, four different state transitions, namely, 0→1, 1→0, 1→1, 0→0 can be achieved through hydrogenation. The four transitions can be reversed through dehydrogenation. The design schematic of QR1 and QR2 is shown in Figure 5b by filling the adopted colors according to the palette in Figure 3 in the corresponding positions before and after hydrogenation, respectively. The detailed sample descriptions can be found in Supporting Figure S4. The experimental results are shown in Figure 5c, which agrees very well with the design schematic in Figure 5b. Before hydrogenation, the microprint shows QR1, presenting information of our website 'www.is.mpg.de/liu'. Upon hydrogen loading, the four aforementioned transitions take place. QR1 is scanned by the hydrogen diffusion fronts and transformed to QR2, presenting information of 'scanning plasmonic color display' (see Supporting Movie 4). Upon oxygen loading, QR2 is switched back to QR1. As a result, dual information can be dynamically encoded within the same plasmonic color microprint. Hydrogen is the only decoding key to read out the protected information, which in the case is QR2. QR2 cannot be decrypted using optical microscopy or even SEM, as QR1 and QR2 are obtained from one single microprint with identical physical features.

   In conclusion, we have demonstrated a novel scheme for the realization of dynamic plasmonic color displays with scanning functions. Such scanning functions are enabled by lateral hydrogenation/dehydrogenation of the microscopic screen through metal/dielectric transitions.

We have shown several examples of scanning plasmonic color displays with different dynamic effects by tailoring the number and geometries of the Pd gates. In addition, we have demonstrated a highly secure encryption approach using a single scanning plasmonic QR code, which contains switchable dual information. Our samples show no significant degradations in switching performance under ambient conditions after three months. For practical applications, gaschromic or electrochromic loading of hydrogen can be utilized for enhanced switching speed and durability.[42–44] Our work advances the established design concepts for plasmonic color printing as well as provides insights into the development of optical information storage and anticounterfeiting features.[45,46]

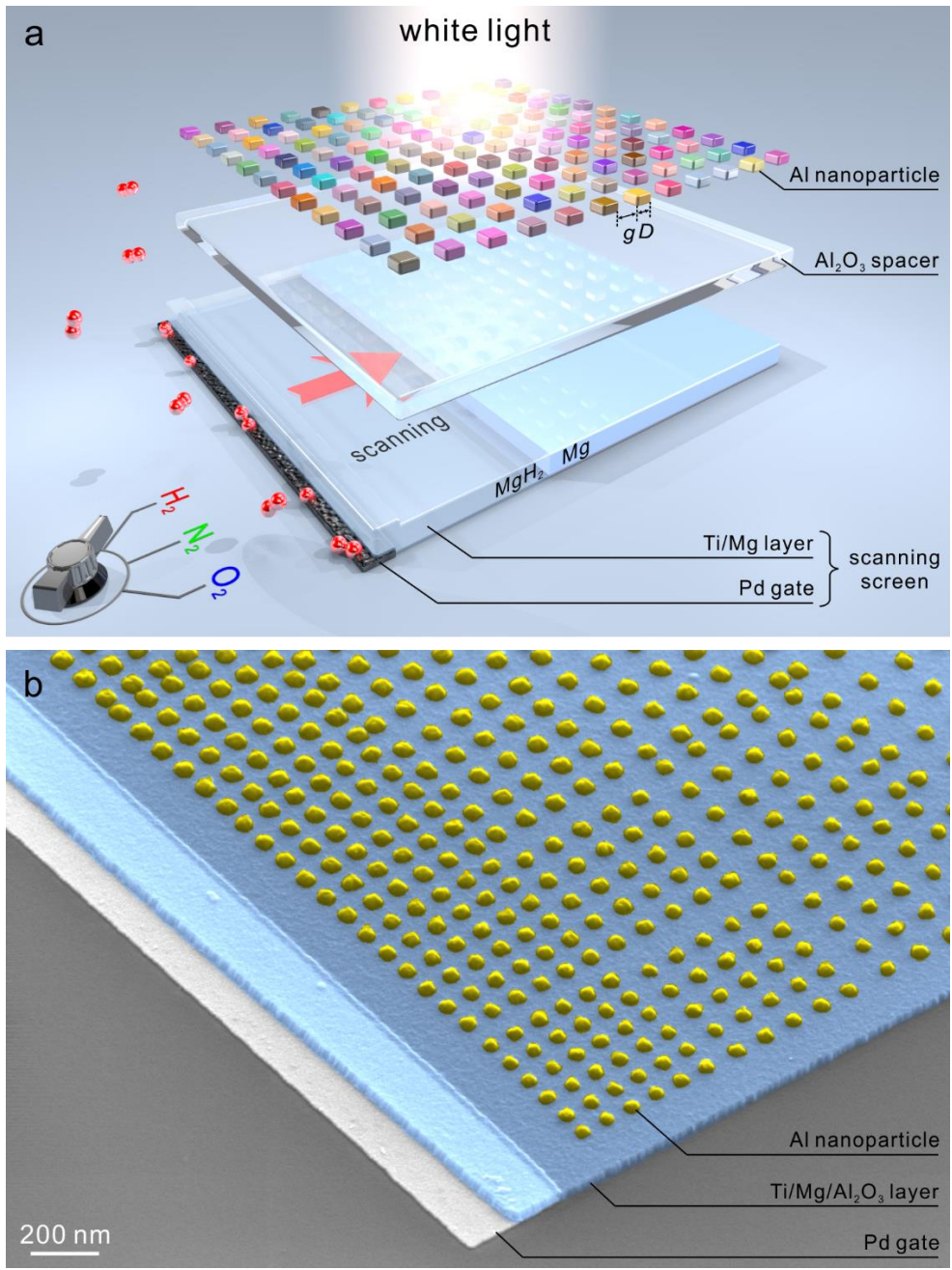

**Figure 1.** (a) Schematic of the scanning plasmonic color display, which consists of Al nanoparticles (diameter *D*; interparticle gap *g*; thickness 25 nm) as plasmonic pixels, a dielectric $Al_2O_3$ (20 nm) spacer, and a scanning Mg screen (15 μm × 15 μm × 30 nm) with a 3 nm Ti buffer layer. The different functional layers are vertically shifted for clarity. The display is illuminated by unpolarized white light. A Pd strip (15 μm × 400 nm × 15 nm) on the left is placed under the edge of the Mg screen. The Pd strip serves as a gate for gas loading or unloading. Upon hydrogen (or oxygen) loading, the Mg to $MgH_2$ (or $MgH_2$ to Mg) transition starts from the Pd gate, giving rise to lateral scanning effects. (b) Tilted SEM image of the left-bottom corner of the scanning plasmonic color display.

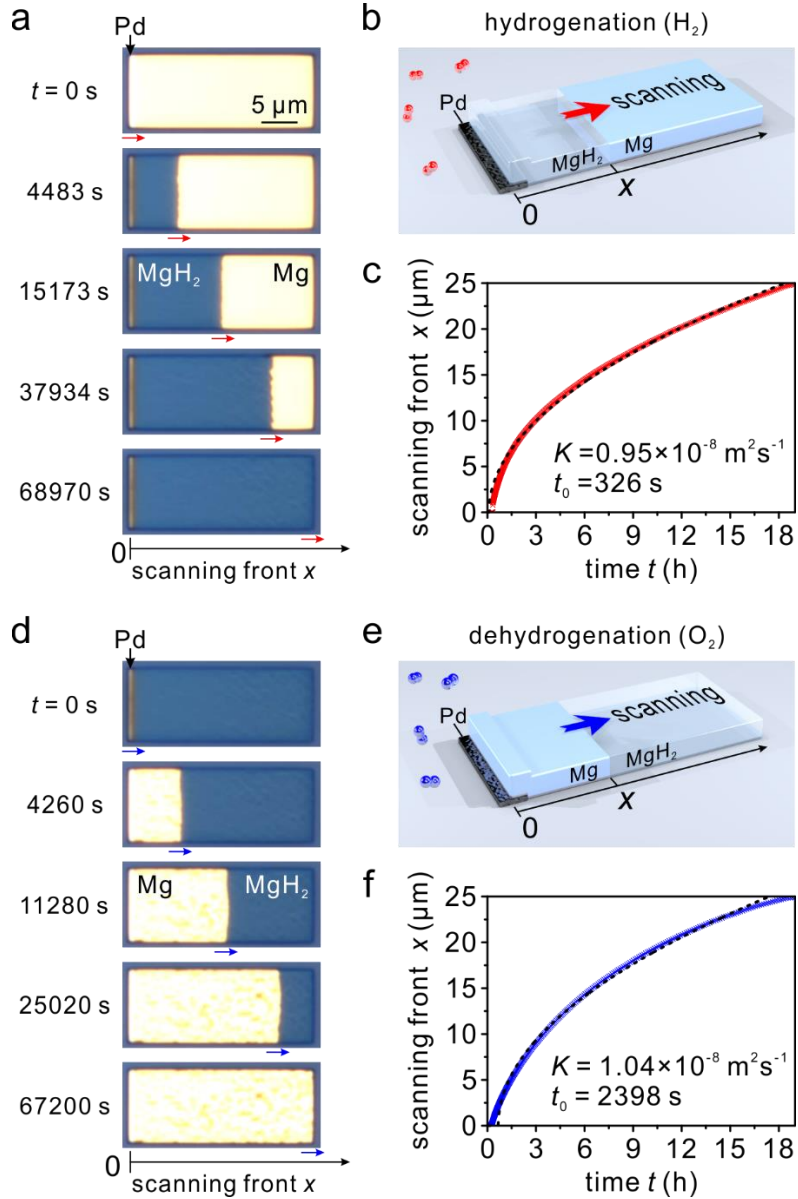

**Figure 2.** Scanning front $x$ of the screen during (a) hydrogenation and (d) dehydrogenation at different diffusion times obtained by *in situ* optical hydrogenography. The red and blue arrows indicate the movements of the diffusion fronts during hydrogenation and dehydrogenation, respectively. The dark and bright regions correspond to $MgH_2$ and Mg, respectively. The sharp color change at the diffusion front is tracked as scanning front $x$. Schematics of the scanning screen during (b) hydrogenation and (e) dehydrogenation. In both cases, scanning starts from the Pd gate. Experimental (red or blue) and fitting (black-dashed) results of the time-dependent scanning front during (c) hydrogenation or (f) dehydrogenation. The fitting parameters $K$ and $t_0$ from Equation (1) are shown for each case.

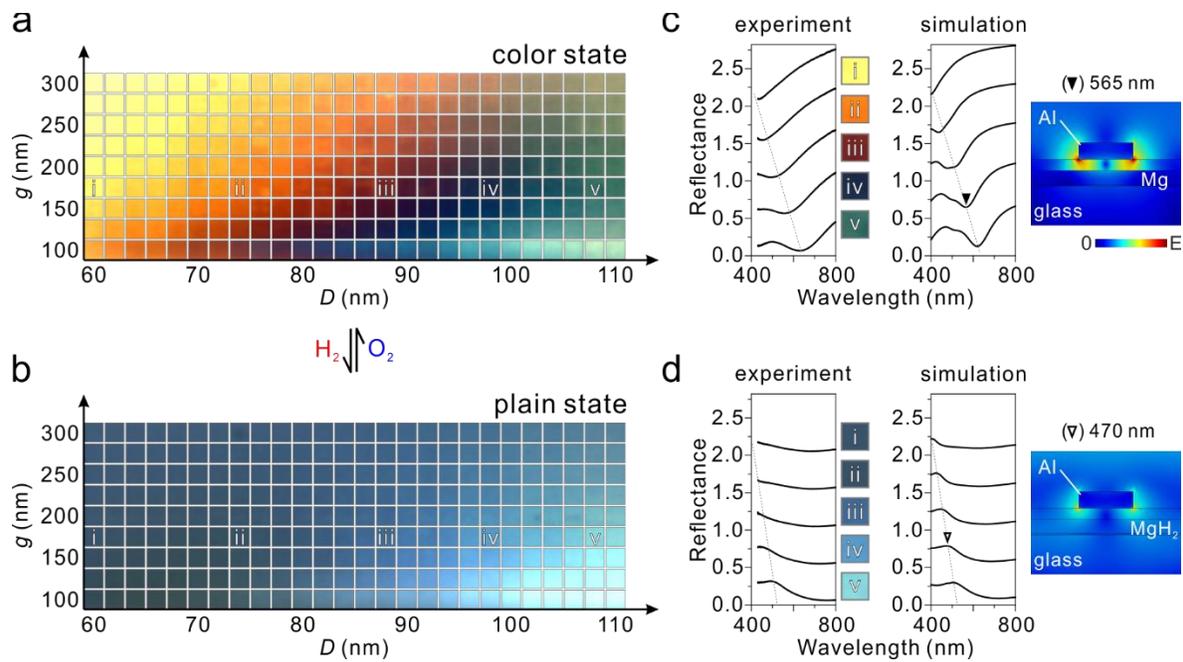

**Figure 3.** Optical micrographs of the palette with stepwise tunning of *D* and *g* at color state (a) and plain state (b). Representative color squares i-v for spectral characterizations are highlighted. The total dimension of each color square is 3 μm. Experimental and simulated reflectance spectra of the selected squares at color state (c) and plain state (d). The spectral curves are shifted upwards stepwise for clarity. The grey-dashed line indicates the shift of the reflectance dip (c) or peak (d). The simulated electric-field distributions of color square v at resonance positions of 565 nm and 470 nm (indicated by arrows) before and after hydrogenation, respectively.

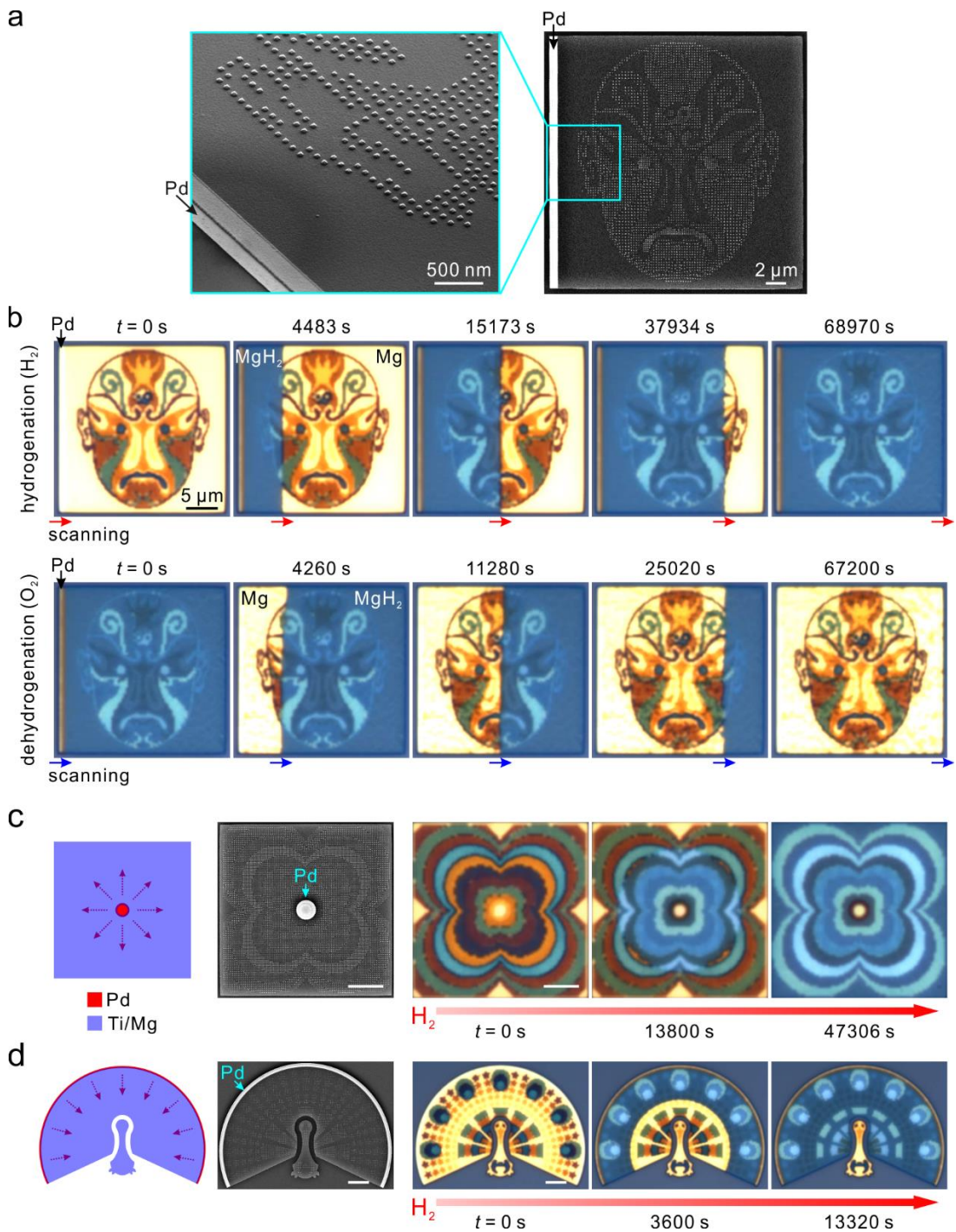

**Figure 4.** (a) Overview and enlarged tilted SEM images of the Sichuan opera facial mask display. (b) Optical micrographs of the display during hydrogenation and dehydrogenation, respectively. The red and blue arrows indicate the scanning directions during hydrogenation and dehydrogenation, respectively. Schematics of the firework (c) and the peacock (d) displays, with the dashed arrows indicating the scanning directions. SEM images of the firework (c) and the peacock (d) displays. Scale bars: 5 μm. Selected snapshots of the firework (c) and the peacock (d) displays during hydrogenation, demonstrating different scanning effects. Scale bars: 5 μm.

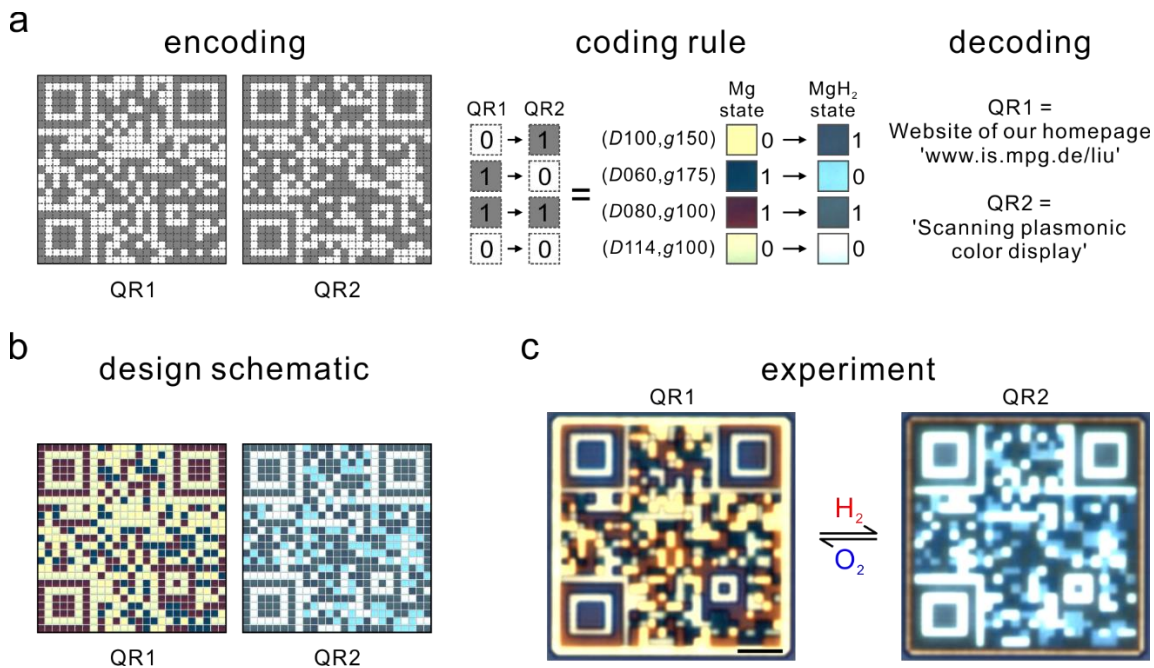

**Figure 5.** (a) Principle of information encoding, coding rule, and decoding. Four different structures are selected from the palette to achieve binary information transitions, namely, '0'→'1', '1'→'0', '1'→'1', and '0'→'0', during hydrogenation. (b) Design schematic of QR1 and QR2 by filling the adopted colors in the corresponding positions. (c) Optical micrographs of the encrypted plasmonic QR code before and after hydrogenation, switching between QR1 and QR2. Scale bar: 5 μm.

AUTHOR INFORMATION

**Corresponding Author**

*E-mail: na.liu@kip.uni-heidelberg.de

Author Contributions

X.D. and N.L. conceived the project. X.D. performed the experiments and theoretical calculations. N.L. wrote the manuscript. Both authors commented on the manuscript.

**Notes**

The authors declare no competing financial interest.

**Acknowledgements**

We gratefully acknowledge the generous support by the 4th Physics Institute at the University of Stuttgart for the usage of clean room facilities. This project was supported by the Sofja Kovalevskaja grant from the Alexander von Humboldt-Foundation, the Marie Curie CIG grant, and the European Research Council (ERC Dynamic Nano) grant.

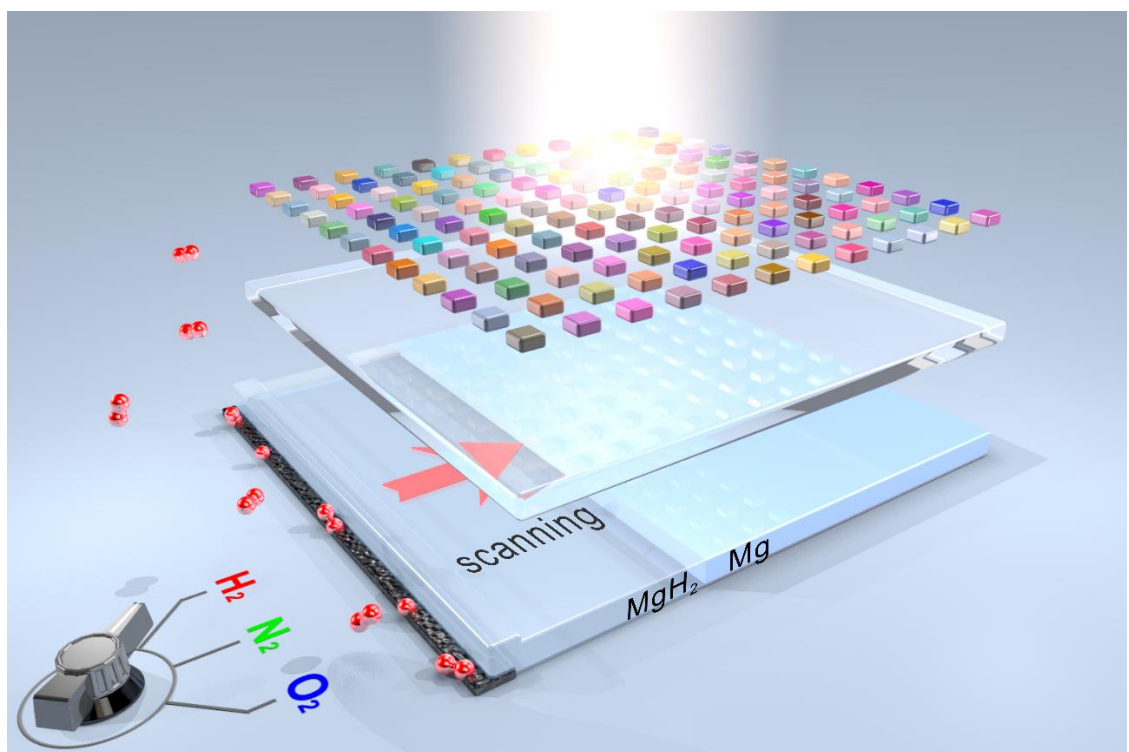

TOC